\documentstyle[]{mn2e}

\def\ni{\noindent}

\footnotesize
\newdimen\minuswidth    
\setbox0=\hbox{$-$}
\minuswidth=\wd0
\catcode`@=\active
\def@{\kern\minuswidth}
\newdimen\digitwidth    
\setbox0=\hbox{\rm0}
\digitwidth=\wd0
\catcode`!=\active
\def!{\kern\digitwidth}
\normalsize

\title[Observations of Fourteen Pulsar Glitches]
{Observations of Fourteen Pulsar Glitches}

\author[A. Krawczyk,  A.G. Lyne, J.A. Gil and B.C. Joshi]
{A.~Krawczyk$^1$, A.G.~Lyne$^2$,  J.A. Gil$^1$ and B.C.~Joshi$^{2,3}$\\
$^1$ Astronomical Centre, Pedagogical University, Lubuska 2, 65-265 Zielona 
G\'ora, Poland\\
$^2$ University of Manchester, Nuffield Radio Astronomy Laboratories, Jodrell
Bank, Macclesfield, Cheshire, SK11 9DL, UK\\
$^3$ National Center for Radio Astrophysics, Pune University Campus, 
P. O. Bag 3, Ganeshkhind, Pune, 411007, India\\}

\begin{document}

\label{firstpage}

\maketitle
\newcommand{\setthebls}{
}
\setthebls

\begin{abstract}
\noindent
About 76 glitches in 25 pulsars have been reported to date. Most glitches
are `giant', with fractional increases of frequency
$\Delta\nu_0/\nu_0\sim 10^{-6}$. 25 glitches were analysed and
presented by Shemar \& Lyne (1996) who detected them mainly at
Jodrell Bank during the monitoring of a sample of 279 pulsars in a
regular timing programme up to MJD 49500. This paper is a continuation 
of their work up to MJD 50500. We present the detection and analysis of 
a further 14 glitches in 9 pulsars, 6 of which have glitched for the 
first time since monitoring had started. Eleven of these glitches are small 
($\Delta\nu_0/\nu_0\sim 10^{-9}$) and below the completeness threshold 
of Shemar and Lyne(1996). We report a giant glitch in PSR B1930+22, the 
second largest reported hitherto, with a $\Delta\nu_0/\nu_0 = 4.5 
\times 10^{-6}$. We also report four recent glitches in PSR B1737$-$30 
which continues to exhibit frequent glitches. Few of these pulsars show 
any recovery after the glitch.
\end{abstract}

\section{Introduction}
Two kinds of irregularities are observed in the rotation rates of
pulsars, timing noise and glitches.  Timing noise is a continuous
wandering of the rotation rate, while glitches are characterized by a
sudden increase in the rate, often followed by a period of relaxation.
They are often revealed by the sudden onset of continually decreasing
arrival time residuals.  Glitches were first observed in the Crab and
Vela pulsars \cite{bgpw69,rm69}
and it was soon realized that they can be important diagnostic tools
for studying neutron star interiors \cite{rud69,bppr69}.
It is widely believed that these events are caused by sudden and
irregular transfer of angular momentum from a faster rotating interior
superfluid to the solid crust of the neutron star. The result is
a sudden fractional increase in the rotational frequency $\nu_0$ of the
pulsar with a magnitude in the range
$10^{-10}<\Delta\nu_0/\nu_0<5\times10^{-6}$. A characteristic feature
of many glitches is a relaxation after the frequency jump, which may occur
over a period of days, months or years. However, as we present in this
paper, small glitches seem to show little significant relaxation after
the glitch.

Although glitches are rather rare phenomena, an increasing number of
these events have been reported recently (McKenna \& Lyne 1990; Shemar
\& Lyne 1996; Wang et al. 2000; this paper)\nocite{ml90,sl96,wmp+00}, 
permitting more comprehensive statistical studies (Alpar \& Baykal 1994; 
McKenna \& Lyne 1990; Lyne, Shemar \& Smith 2000)\nocite{ab94,ml90,lsg00}.  
In this paper, we extend the analysis of the Jodrell Bank database by 
2.5~year 
beyond that of Shemar \& Lyne (1996) and seek to lower the detection
threshold significantly.

\section{Observations and analysis}
Observations were carried out at Jodrell Bank, mostly using the 76-m
Lovell radio telescope, but also occasionally using the 30-m Mark II
telescope. Measurements were made at intervals of between one and
three months, while some, more interesting, pulsars were observed more
often. The total list of pulsars that have been observed regularly in
this programme at Jodrell Bank is presented in Shemar \& Lyne (1996,
their Table 1)\nocite{sl96}.  Those authors analysed the data up to
about MJD 49500. This work represents a more detailed study of the
same data and also extends the analysis on their list of pulsars to
about MJD 50500. The B1950 and J2000 names of these pulsars, their periods 
and characteristic ages, and the dates spanned by the observations are 
listed in Table 1. The main improvement in the analysis is a more
careful calibration of the systematic effects which arose from the use
of a variety of filterbanks and dedispersion procedures over the
typically 16-year span of the observations.  Greater care was also
exercised in removing data which might have been affected by impulsive
radio-frequency interference.

Both telescopes were equipped with dual-channel cryogenic receivers at
observing frequencies centered close to 408, 610 or 1400 MHz.  Each
channel was sensitive to one hand of circular polarization. The data
were dedispersed using filterbanks and folded synchronously with the
nominal topocentric rotation period of the pulsar for sub-integration
periods of between one and three minutes.  An observation consisted
typically of six such integrations which were stored on disk for
subsequent processing.

{\small
\begin{table*}
\begin{center}
\begin{tabular}{lclrcr}
\multicolumn{6}{l}{\large{\bf Table 1.} Data span of Timing Observations used                          for analysis}  \\
\\
\hline
    PSR J   &  PSR B  &     Period    &  Age  & MJD   RANGE   &  NO OF TOA \\
            &         &       (s)     & (Kyr) &               &     \\ \hline
  0157+6212 & 0154+61 & 2.35172383222 &  200  & 46866 $-$ 50496 &  231  \\
  1740$-$3015 & 1737$-$30 & 0.60666591713 &   20  & 49243 $-$ 51300 &  223  \\
  1801$-$0357 & 1758$-$03 & 0.92148958467 & 4400  & 46718 $-$ 50586 &  124  \\
  1801$-$2304 & 1758$-$23 & 0.41579643949 &   60  & 49700 $-$ 50687 &   71  \\
  1803$-$2137 & 1800$-$21 & 0.133634078897&   16  & 49403 $-$ 50600 &  127  \\
  1910$-$0309 & 1907$-$03 & 0.50460431337 & 3600  & 47392 $-$ 50530 &   92  \\
  1919+0021 & 1917+00 & 1.27225573197 & 2600  & 48104 $-$ 50640 &    175  \\
  1932+2220 & 1930+22 & 0.144455311469&   40  & 49402 $-$ 50583 &    112  \\
  2257+5909 & 2255+58 & 0.368245626351& 1000  & 47523 $-$ 50588 &    134 \\
\hline
\end{tabular}
\end{center}
\end{table*}
}

{\small
\begin{table*}
\begin{center}
\begin{tabular}{llll}
\multicolumn{4}{l}{\large{\bf Table 2.} Assumed Positions of 9 Glitching Pulsars} \\
\\
\hline
\multicolumn{1}{c}{NAME} 
 & \multicolumn{1}{c}{RA(J2000)} & \multicolumn{1}{c}{Dec(J2000)} 
 & \multicolumn{1}{l}{Reference} \\
\hline
B0154+61  & 01 57 49.91 & +62 12 25.328 & Martin (2001)\\
B1737$-$30 & 17 40 33.82(1) & $-$30 15 43.5(2) & Fomalont et al. (1997) \\
B1758$-$03 & 18 01 22.66 & $-$03 57 55.39 & Martin (2001)\\ 
B1758$-$23 & 18 01 19.803(9) & $-$23 04 44.2(2) & Frail et al. (1993) \\
B1800$-$21 & 18 03 51.401(4) & $-$21 37 07.34(7) & Fomalont et al. (1992)  \\
B1907$-$03 & 19 10 29.686 & $-$03 09 54.318 & Martin (2001)\\
B1917+00 & 19 19 50.654 & +00 21 39.848 & Martin (2001)\\
B1930+22 & 19 32 22.693 & +22 20 53.68  & This paper \\
B2255+58 & 22 57 57.741 & +59 09 14.917 & Martin (2001)\\ 
\hline
\end{tabular}
\end{center}
\end{table*}
}

Total intensity profiles were obtained by adding the six
sub-integrations. These were then cross-correlated with a standard
template to give pulse topocentric times of arrival which were
then corrected to the Solar system barycentre using the JPL ephemeris
DE200 \cite{sta82}.  Assessment of arrival time residuals, which are
the differences between actual pulse arrival times and times
calculated from a simple rotational model, provides information about
the behaviour of the pulsar rotation.  The fitting procedure used a
simple slow-down model involving rotational frequency and its first
derivative. The analysis for each pulsar involves such a fit to a
period of time which is devoid of any glitch activity.  The timing
residuals for the whole data set are then inspected visually for the
presence of glitches.  Pulsar positions assumed in this analysis are
given in Table 2\nocite{fgl+92,fkv93}.  Several of the positions were
obtained from the same timing data described here, using data located
well away from glitches (Martin 2001\nocite{mar01}).

The parameters of each glitch were obtained from comparison of
parameters before and after the glitch.  Epochs of glitches were
determined by requiring a continuity of phase across the glitch. 
Pre-glitch
parameters were obtained by fitting a simple pre-glitch slow-down
model of the form $\nu(t) = \nu_0 + \dot{\nu}_0t$ to the 
data. The observed post-glitch frequency residuals are described as a
function of the time $t$ elapsed since the epoch of the glitch,
relative to the pre-glitch ephemeris:
\begin{equation}
\Delta\nu(t) =
\Delta\nu_p + \Delta\dot{\nu}_p t+\ddot{\nu}_p t^2/2 +
\Delta \nu_1e^{-t/\tau_1},
\end{equation}
where $\Delta\nu_p = \nu_p -
\nu_0$ and $\Delta\dot{\nu}_p = \dot{\nu}_p - \dot{\nu}_0$ are
differences between post-glitch and pre-glitch parameters. 
The last
term in equation (1) describes an exponentially decaying transient
component of post-glitch behaviour, while the penultimate term
represents the large, approximately constant value of second
derivative often seen following large glitches after any short-term
transient has decayed. 
Note that $\Delta\nu_{p}$ and $\Delta\dot{\nu}_{p}$ usually differ from the
instantaneous changes in $\Delta\nu_{o}$ and its derivative because of
the decaying components. Thus,
\begin{equation}
\Delta\nu_{p} = \Delta\nu_{o} - \Sigma \Delta\nu_{1}
\;\;{\rm and}\;\;
\Delta\dot{\nu}_{p} = \Delta\dot{\nu}_{o} + \Sigma \Delta\nu_{1}/\tau_{1} 
\end{equation}
A more detailed description of the
observation system and analysis can be found in the paper by Shemar \&
Lyne (1996).

\section{Results}
The parameters of 14 new glitches found in 9 pulsars  
\nocite{wmp+00} 
are given in
Table 3, which shows the epoch of the glitch as a Modified Julian Date
(MJD), pre-glitch frequency $\nu_0$ and its first derivative
$\dot{\nu}_0$ at that epoch, glitch fractional parameters, and the
post-glitch frequency $\nu_p$ and its first derivative $\dot{\nu}_p$.
The glitch epoch is estimated in the following manner. First, two solutions 
for the pulse phase across the glitch were obtained by extrapolating the 
pulse ephemeris before and after the glitch respectively. Then, the epoch 
was estimated from these phases by requiring that the the pulse 
phase be continuous across the glitch.  The errors quoted in
brackets are twice the standard deviations obtained from the formal
fits. However, this procedure could not be used for the glitch in 
PSR B1930+22 due to lack of sufficient number of measurements near the 
glitch as well as the large magnitude of the glitch. 
The relaxation parameters, $\dot{\nu}_p$, $\Delta \nu_1$ and
$\tau_1$ are usually insignificant in these glitches and are mentioned
in the text where appropriate.

{\small
\begin{table*}
\begin{center}
\begin{tabular}{llllllllll}
\multicolumn{10}{l}{\large{\bf Table 3.} Pre-glitch, glitch and post-glitch parameters
for 14 glitches in 9 pulsars. Errors in the least significant}\\
\multicolumn{10}{l}{\large place are given in parentheses.}\\
\\
\hline
 & & \multicolumn{2}{c}{Pre-glitch Parameters} && \multicolumn{2}{c}{Glitch Parameters} && \multicolumn{2}{c}{Post-glitch Parameters}\\
\cline{3-4} \cline{6-7} \cline{9-10}
PSR B & Epoch & \multicolumn{1}{c}{$\nu_0$} & \multicolumn{1}{c}{$\dot{\nu}_0$} && $\Delta\nu_0/\nu_0$ 
       & $\Delta\dot{\nu}_0/\dot{\nu}_0$ && \multicolumn{1}{c}{$\nu_p$} & \multicolumn{1}{c}{$\dot{\nu}_p$} \\
      & (MJD) & \multicolumn{1}{c}{$(s^{-1})$} & \multicolumn{1}{c}{$(10^{-15} s^{-2})$} && $(10^{-9})$  
       & $(10^{-3})$ && \multicolumn{1}{c}{$(s^{-1})$} & \multicolumn{1}{c}{$(10^{-15} s^{-2})$} \\
\hline \vspace{2mm} 
0154+61 & 48504(1) & 0.42521973638(2) & !!$-$34.1638(3) && !!!2.46(6) & $-$0.04(1) && 0.42521973743(1) & 
!!$-$34.1625(1) \vspace{2mm} \\ 
1737$-$30 & 49451.7(4)  & 1.6483288501(3) & $-$1265.76(2)&& !!!9.5(5) & $-$0.32(2) && 1.6483288657(7) & 
$-$1265.4(2)  \vspace{2mm} \\ 
          & 49543.93(8) & 1.6483203659(8) & $-$1265.3(2) && !!!3.0(6) & $-$0.68(2) && 1.6483203709(6) & 
$-$1264.48(1) \vspace{2mm} \\
          & 50574.5497(4) & 1.6482078436(2) & $-$1264.02(1) && !439.3(2) & @1.261(2) && 1.6482085677(2) & 
$-$1265.62(2) \vspace{2mm} \\ 
          & 50941.6182(2) & 1.6481684365(2) & $-$1265.56(1) && 1443.0(3) & @1.231(5) && 1.6481708149(5) & 
$-$1267.12(6) \vspace{2mm} \\
1758$-$03 & 48016(4) & 1.0851988198(2) & !!!$-$3.899(3) && !!!2.9(2)  & @1.17(9) && 1.08519882303(2) &
!!!$-$3.903(1) \vspace{2mm}  \\ 
1758$-$23 & 50055.0(4) & 2.405065322(2) &  !$-$653.5(3) &&   !!22.6(9) & $-$0.08(2) && 2.405065377(1) &  
!$-$653.43(9) \vspace{2mm} \\ 
          & 50363.414(4) & 2.405047996(1) & !$-$653.42(9) &&   !!80.6(6) &  @0.50(2) && 2.4050481894(9) &
  !$-$653.75(6) \vspace{2mm} \\ 
1800$-$21 & 50269.4(1) & 7.483583400(1) & $-$7496.36(8) && !!!5.3(2) & @0.195(4) && 7.483583440(2) & 
$-$7497.8(3) \vspace{2mm} \\ 
1907$-$03 & 48241(2) & 1.9817509328(1) & !!$-$8.600(3) && !!!0.60(6) & @1.04(4) && 1.98175093394(6) & 
!!$-$8.609(1) \vspace{2mm} \\ 
    & 49219.85(2) & 1.98175020655(9) & !!!$-$8.606(2) && !!!1.84(6) & @0.28(3)&& 1.98175021019(6) & 
!!!$-$8.609(1) \vspace{2mm} \\ 
1917+00 & 50174(2) & 0.78600232349(1) & !!!$-$4.741(1) && !!!1.29(3) & @0.559(9) && 0.78600232450(2) &
!!!$-$4.744(4) \vspace{2mm} \\ 
1930+22 & 50264(20) & 6.92210791(2) & $-$2756.4(1) && 4457(6) & !1.7(2) && 6.92213877(2) &
$-$2761.0(3)  \vspace{2mm} \\
2255+58 & 49463.2(2) & 2.71557492794(4) & !!$-$42.436(1) && !!!0.92(2) & $-$0.032(2) && 2.71557493043(4) &
!!$-$42.434(1) \vspace{2mm} \\
\hline
\end{tabular}
\end{center}
\end{table*}
}


The frequency residuals for previously unpublished glitches are
presented in the lower panels of Figs.~1, 3, 5$-$9 and Fig.~11, and in the
top panel of Fig.~10. They were obtained by performing local fits over
about 50 days to the arrival time data, and presented relative to a
simple slow-down model.  Since most of the glitches presented in this
paper are small, we also usually show their timing residuals in the
upper panels of Figs.~1, 3, 5$-$9 and Fig.~11 for clarity of
presentation.  The number of available timing measurements between two 
glitches was small in the case of PSR B1737$-$30 making it difficult 
to obtain the frequency residuals in the manner described above except 
in case of two glitches. The timing residuals in all these pulsars 
show the familiar negative change in the gradient after the glitch,  
corresponding to a spin-up.

Below we describe detailed results of the search for glitches in the
improved Jodrell Bank pulsar timing data base. Following Shemar \&
Lyne (1996), we give for each pulsar the conventional B-name, J-name
\cite{tml93} and the characteristic age $\tau =
-\nu/2\dot{\nu}$.

\begin{figure}
\setlength{\unitlength}{1cm}
\begin{picture}(8,16)
\put(-1.7,-1.5){\includegraphics{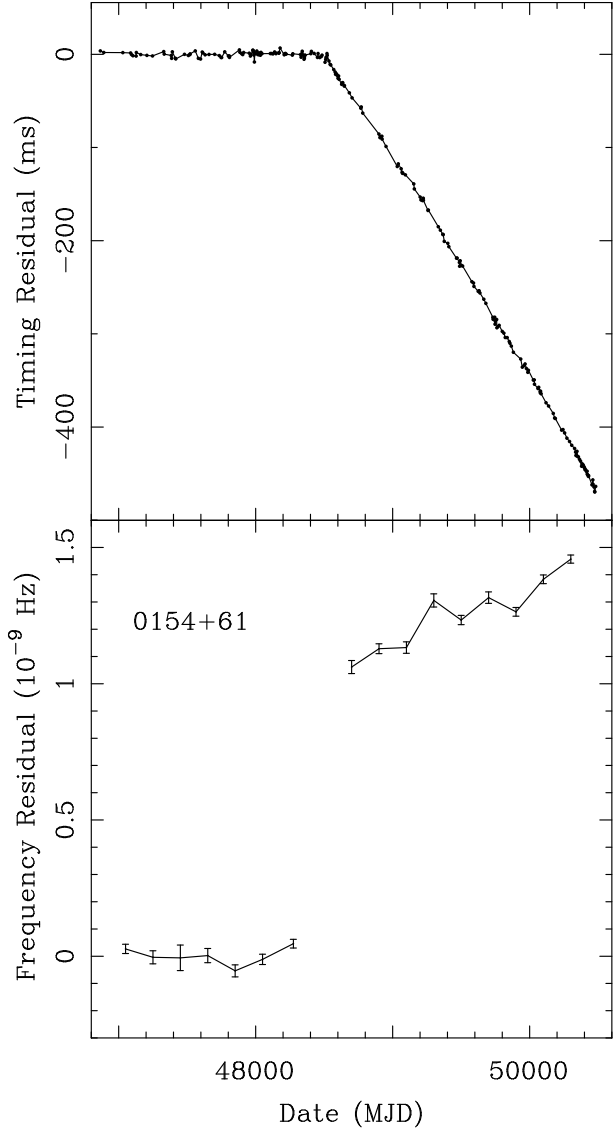}}
\end{picture}
\caption{Rotational history of PSR~B0154+61 with timing and frequency residuals
presented in upper and lower panels, respectively. Errors are smaller than 
size of the points in the upper panel and are typically 0.6 ms}
\label{fig1}
\vspace{0.7in}
\end{figure}

\begin{figure}
\setlength{\unitlength}{1cm}
\begin{picture}(8,16)
\put(-1.7,-1.5){\includegraphics{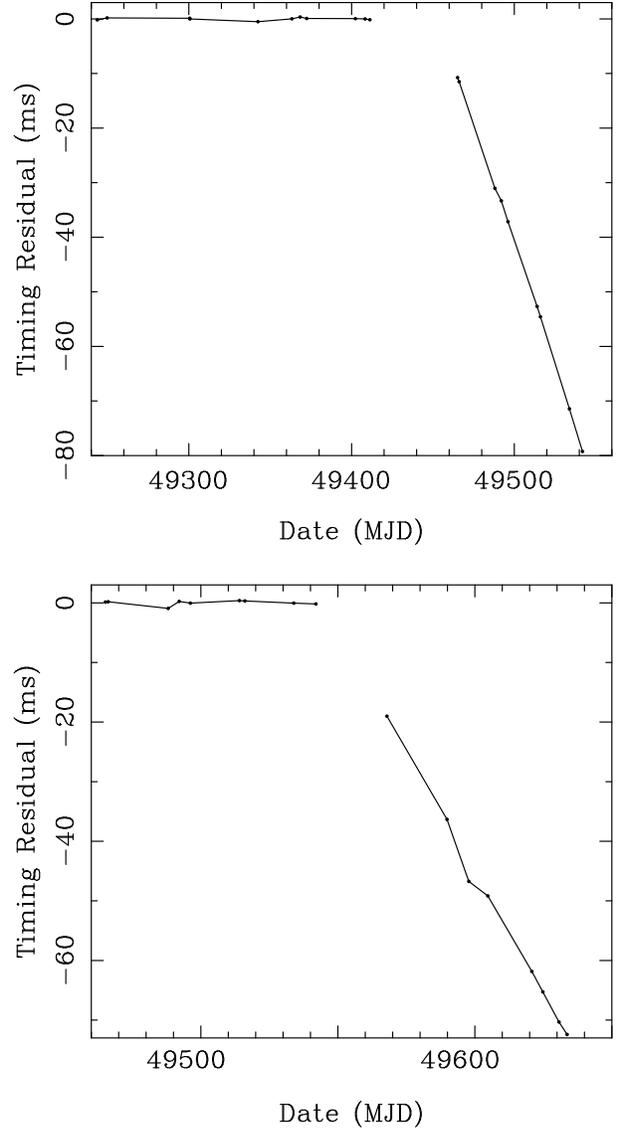}}
\end{picture}
\caption{Timing residuals for the two small glitches in PSR B1737$-$30. The 
residuals for the glitch at MJD 49451 are presented in upper panel, while 
those for the glitch at MJD 49544 are shown in the lower panel, respectively.
Errors are smaller than size of the points and are typically 0.2 ms}
\label{fig2}
\vspace{2.0in}
\end{figure}

\subsection{PSR~B0154+61 (J0157+6212, $\tau=200$~kyr)}
Although data on PSR~B0154+61 have been collected at Jodrell Bank for
more than 10 years since MJD 46866, this pulsar suffered its first
observed glitch near MJD 48504. The glitch was quite small
with the size of the frequency jump equal to $2.46 \times
10^{-9}$. The frequency and timing residuals are shown in lower and
upper panels of Fig.~1, respectively. This glitch was not reported by
Shemar \& Lyne (1996), as it was below their completeness level of $5
\times 10^{-9}$ in $\Delta\nu_0/\nu_0$.

\subsection{PSR~B1737$-$30 (J1740$-$3015, $\tau=20.63$~kyr)}

This pulsar exhibits frequent glitches and nine glitches have been reported 
in the past (McKenna \& Lyne 1990; Shemar and Lyne 1996)\nocite{ml90}. Our 
analysis included observations carried out up to May 1999 for this pulsar 
(around 51300 MJD). We present four more glitches detected in these data. 
The timing residuals for the two smaller glitches are shown in Figure 2. 
The other two glitches were large with a $\Delta\nu_0/\nu_0$ exceeding 
$1.0 \times 10^{-7}$.  The timing and frequency residuals for these 
glitches are shown in Figs. 3a and 3b and the glitch parameters for all 
the four glitches are presented in Table 3. 

The cumulative fractional change in the rotation rate, 
$\Delta\nu_0/\nu_0$, for the pulsar is shown in Fig. 4. The dashed - dot 
line indicates the average rate of fractional spin-up due to glitches. The 
mean spin-up rate due to glitches, $\dot{\nu}_{glitch}$ 
(See Lyne et al. 2000), for this pulsar is about 1.4 percent of its 
spin-down rate. Thus, a fixed fraction 0.014 of the pulsar's slowdown is 
reversed by glitch activity and this is consistent with statistical 
estimate in other pulsars (Lyne et al. 2000). Shemar and Lyne 
(1996) noted that there are two types of glitches and this is 
evident from this figure. The larger glitches occur typically 800 days 
apart whereas the typical separation for all the glitches is of the 
order of 300 days.

\subsection{PSR~B1758$-$03 (J1801$-$0357, $\tau=4,400$~kyr)}

The rotational frequency and timing residuals for this pulsar are
shown in Fig. 5.  It suffered a glitch after 3 years of regular
monitoring at Jodrell Bank and was not reported by Shemar \& Lyne
(1996), being below their sensitivity threshold.  The size of the
frequency jump is rather small, with the fractional increase
$\Delta\nu_0/\nu_0 \approx 3 \times 10^{-9}$.  The only older pulsar
which has been observed to glitch is PSR~B1859+07 ($\tau = 4.5$ Myr).
It is difficult to determine the exact date of the glitch because of
the large gap of 140 days between observations. An estimate of the glitch 
epoch by assuming the continuity of pulse phase across the glitch indicates 
that the glitch occurred sometime near the beginning of 1990 (around 
MJD 48016). It was probably followed by a small relaxation, visible 
in the lower panel of Fig. 5.

\vspace{1.0in}

{\begin{figure*}
\setlength{\unitlength}{1cm}
\begin{picture}(16,16)
\put(-1.7,-1.5){\includegraphics{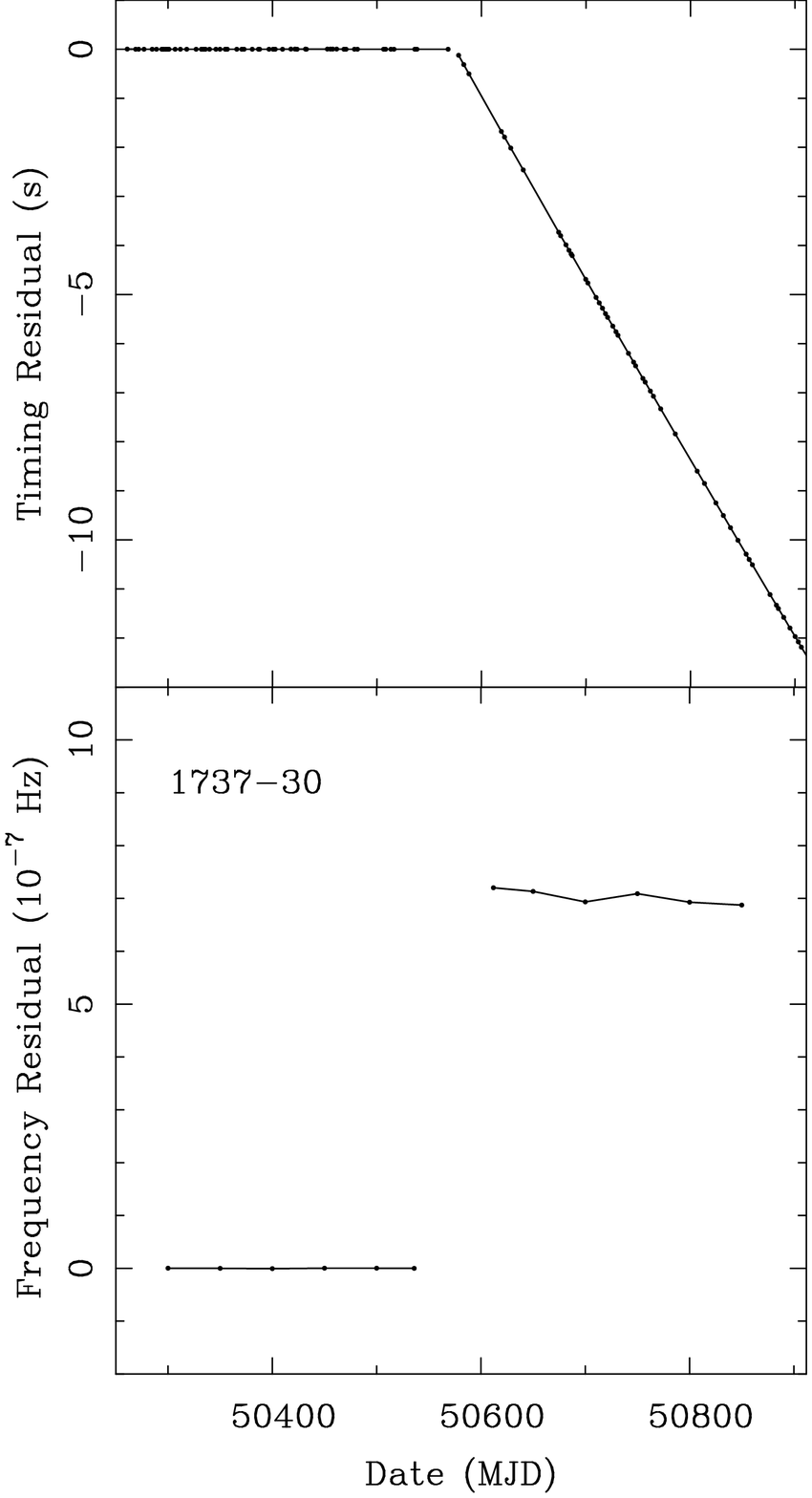}}
\put(7.0,-1.5){\includegraphics{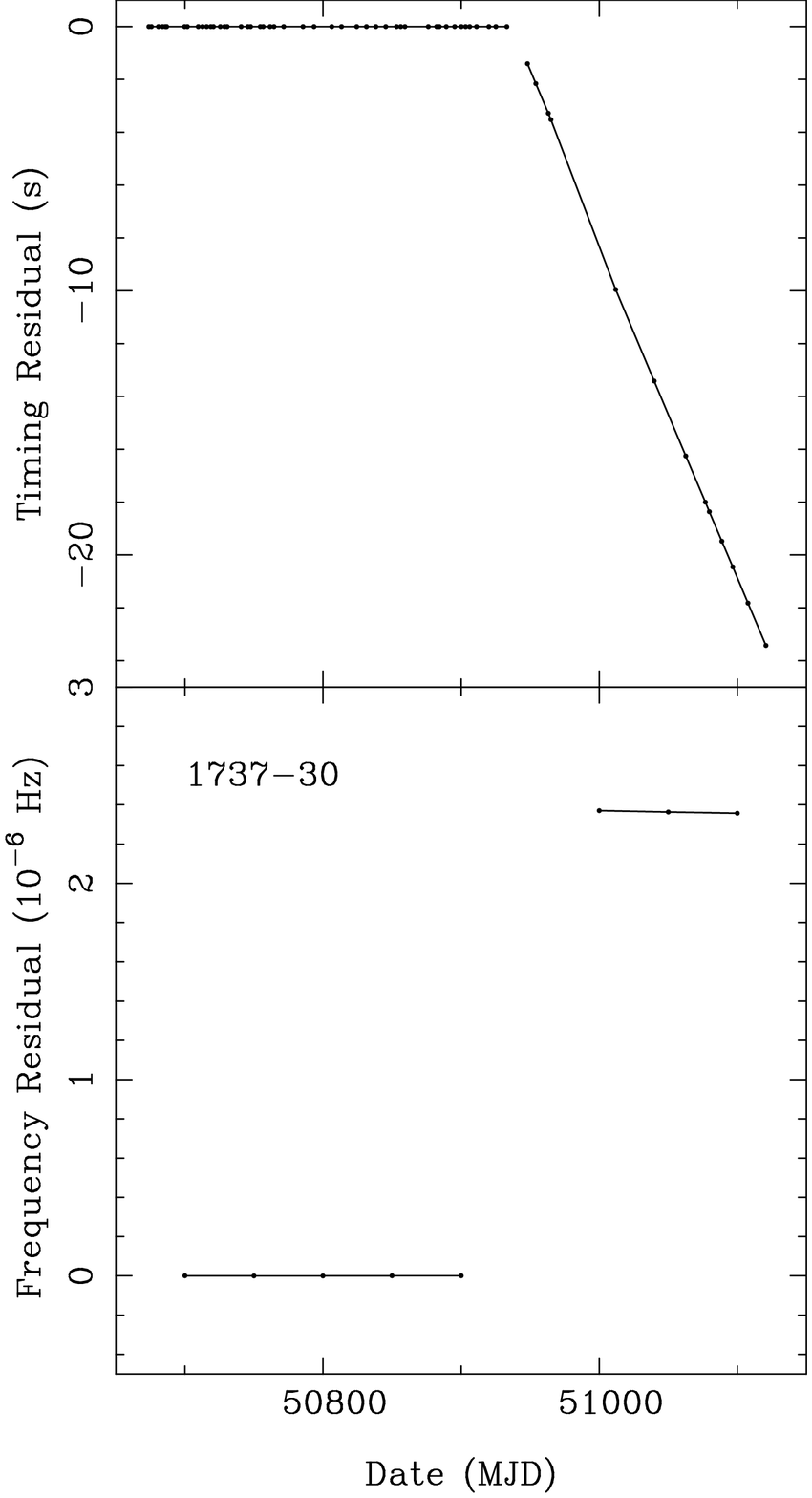}}
\end{picture}
\caption{Rotational history of PSR~B1737$-$30 with timing and frequency 
residuals presented in upper and lower panels, respectively. The glitch 
around MJD 50574 is shown in (a) and that around MJD 50941 in (b). Errors 
are smaller than size of the points in all plots and are typically 0.1 ms 
in the upper panel and of the order of $10^{-9}$ Hz in the lower panel}
\label{fig3}
\end{figure*}
}

\subsection{ PSR~B1758$-$23 (J1801$-$2306, $\tau=60$~kyr)}

The frequency and timing residuals for this pulsar are shown in
Fig. 6.  A total of 6 glitches have been detected in PSR~B1758$-$23
since MJD 46697.  The first three glitches were reported by Kaspi et
al. (1993)\nocite{klm+93}, the next one by Shemar and Lyne
(1996)\nocite{sl96} and we show two more recent glitches here. Four of 
these glitches were also reported by Wang et al. (2000). The
first occurred around MJD 50055 and the second about a year
later. Both newly-reported glitches are of moderate size, having
magnitudes of $\Delta\nu_0/\nu_0 = 22.6
\times 10^{-9}$ and $80.6
\times 10^{-9}$, respectively. These values are consistent with those 
reported by Wang et al. (2000). The event reported in Shemar \& Lyne
(1996) is of similar size to the events presented here, while the
glitches recorded by Kaspi et al. (1993)\nocite{klm+93} are one order
of magnitude larger, but still substantially less than those seen in
Vela and other youthful pulsars. Thus all glitches in this pulsar are
rather small. The post-glitch data analysis does not show any
significant relaxation in this pulsar.

\vspace{1.0in}
 
\begin{figure}
\setlength{\unitlength}{1cm}
\begin{picture}(10,10)
\put(-1.7,-1.5){\includegraphics{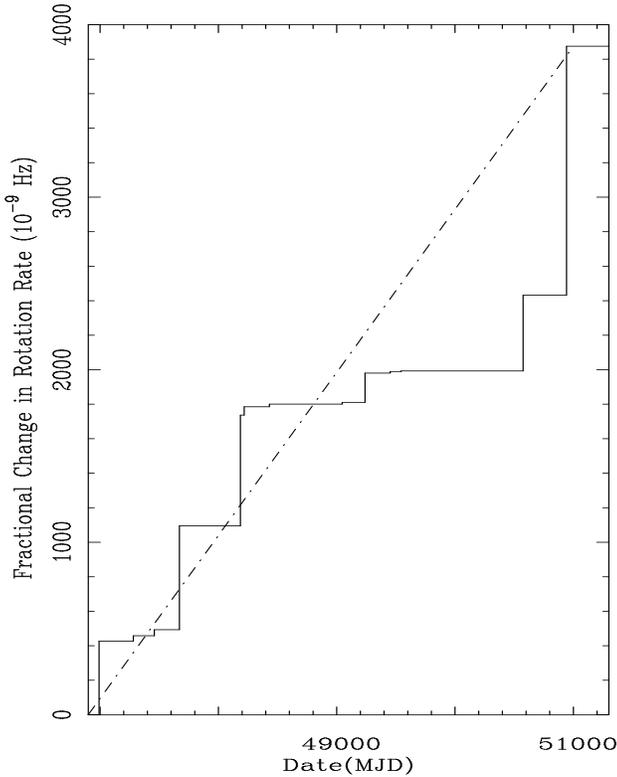}}
\end{picture}
\caption{Cumulative fractional change, $\Delta\nu_0/\nu_0$, in the rotation 
rate of PSR~B1737$-$30 for all the reported glitches. The average rate of 
fractional spin-up is indicated by the dashed line}
\label{fig4}
\vspace{1.5in}
\end{figure}

\begin{figure}
\setlength{\unitlength}{1cm}
\begin{picture}(8,16)
\put(-1.7,-1.5){\includegraphics{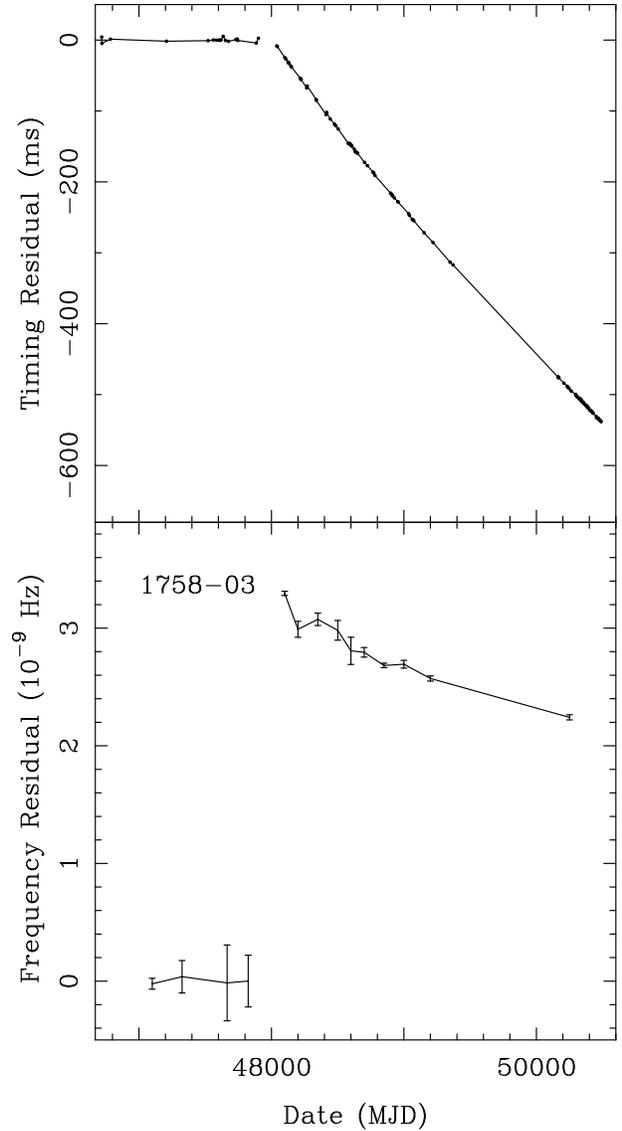}}
\end{picture}
\caption{Rotational history of PSR~B1758$-$03 with timing and frequency 
residuals presented in upper and lower panels, respectively. Errors are 
smaller than size of the points in the upper panel and are typically 0.4 ms}
\label{fig5}
\vspace{1.0in}
\end{figure}

\subsection{PSR~B1800$-$21 (J1803$-$2137, $\tau=16$~kyr)}

Shemar \& Lyne (1996) reported a giant glitch with the third largest
magnitude of all at the end of 1990, while we have found a small glitch
with $\Delta\nu_0/\nu_0 = 5.3 \times 10^{-9}$ that occurred more
than five years later around MJD 50269. The relatively dense data
coverage around the event permitted the determination of the time of
the event to within about one day.  The timing and frequency
residuals are presented in Fig. 7 (relative to data from about 1000
days between glitches).  These plots show that the glitch was
accompanied by a significant increase in the rate of slow-down.

\vspace{0.3in}

\begin{figure*}
\setlength{\unitlength}{1cm}
\begin{picture}(16,16)
\put(-1.7,-1.5){\includegraphics{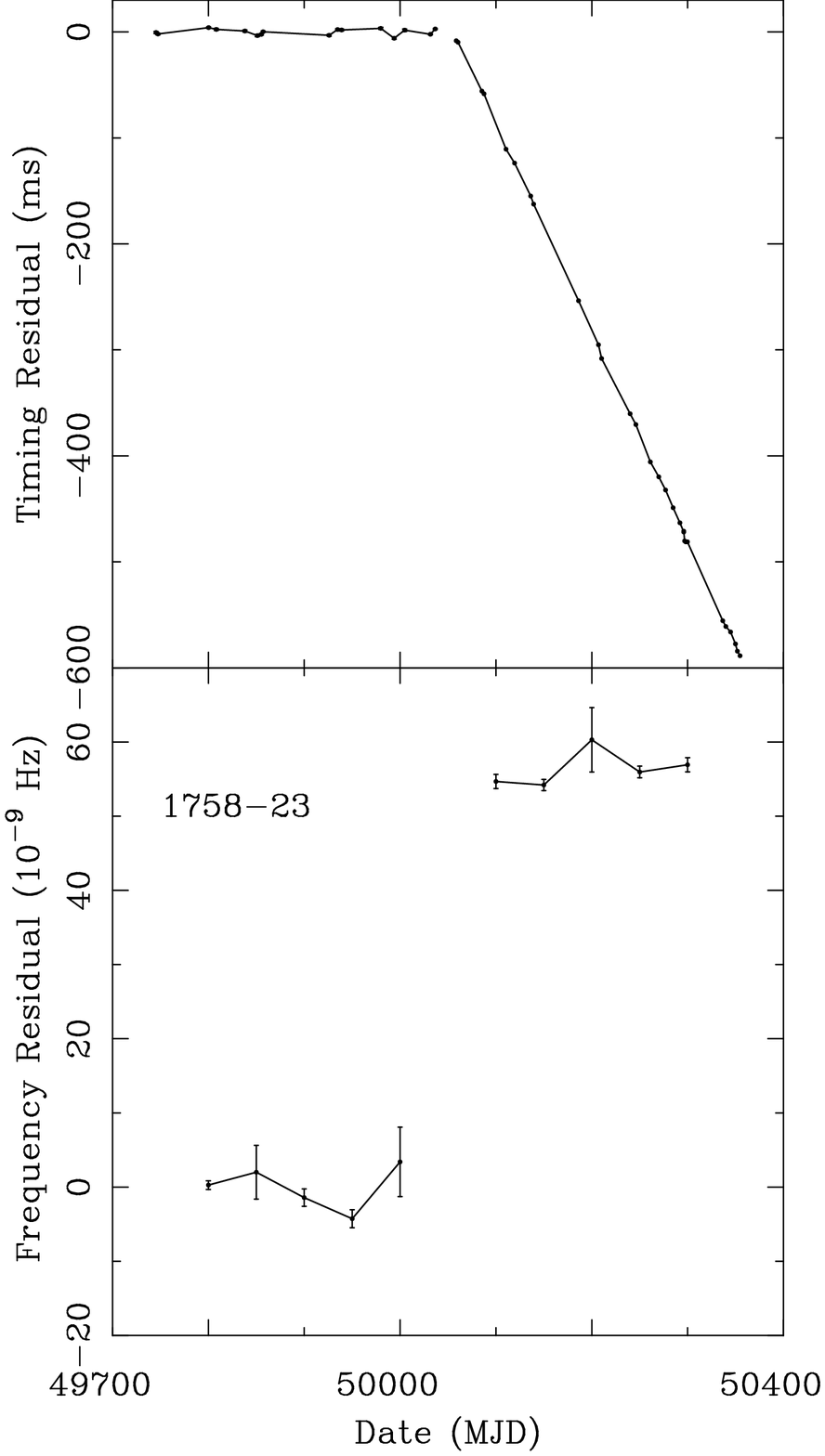}}
\put(7.0,-1.5){\includegraphics{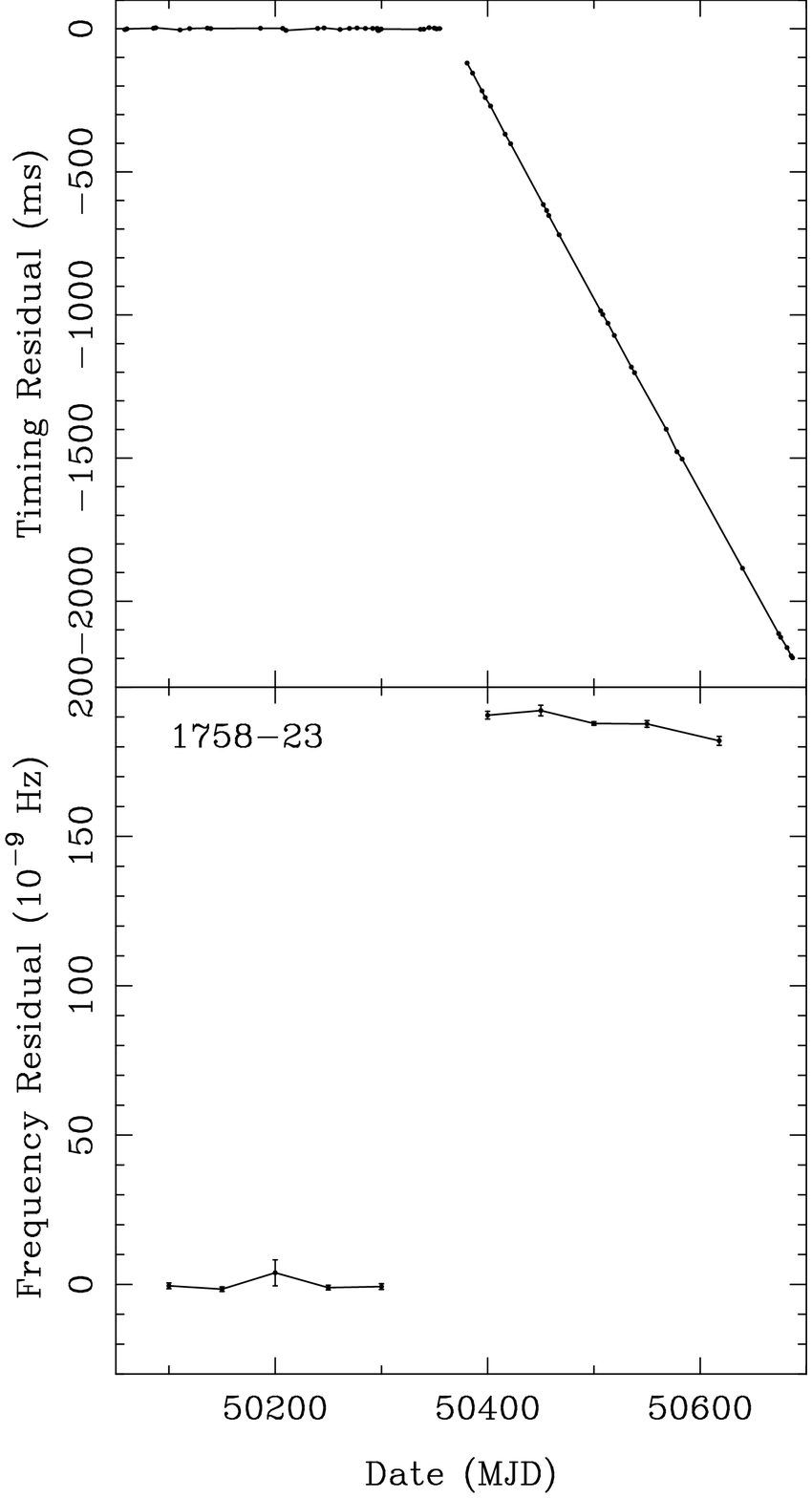}}
\end{picture}
\caption{Rotational history of PSR~B1758$-$23 with timing and frequency 
residuals presented in upper and lower panels, respectively. The glitch 
around MJD 50055 is shown in (a) and that around MJD 50363 in (b). Errors 
are smaller than size of the points in the upper panels and are typically 
0.1 ms }
\label{fig4q}
\vspace{2.0in}
\end{figure*}

\subsection{PSR~B1907$-$03 (J1910$-$0309, $\tau=3,600$~kyr)}
This is one of the oldest pulsars known to glitch and it suffered two
rather small glitches. One of them occurred around MJD 48241 and the
other about 1000 days later. In Fig. 8a and 8b, one can see the timing and 
frequency residuals for this pulsar corresponding to the two glitches. 
The fractional frequency increases are 
$0.6 \times 10^{-9}$ and $1.84 \times 10^{-9}$, respectively. The
first glitch is the smallest glitch known.  Again,
timing noise is probably the reason for irregular behaviour both
before and after glitches. Both glitches reported here were below the
threshold of the search of Shemar \& Lyne (1996).

\begin{figure}
\setlength{\unitlength}{1cm}
\begin{picture}(8,16)
\put(-1.7,-1.5){\includegraphics{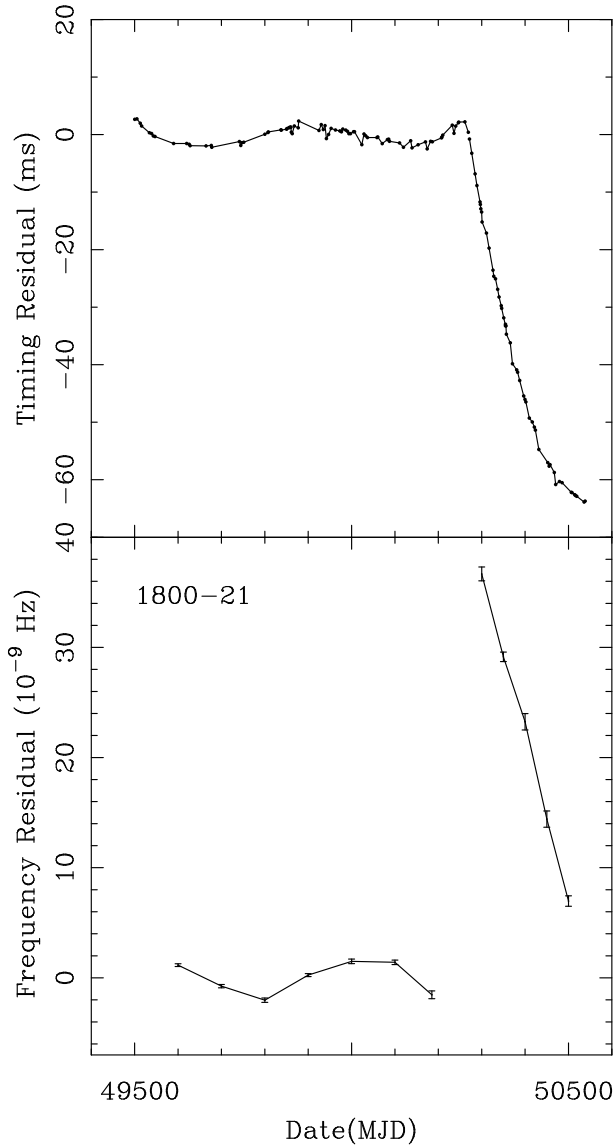}}
\end{picture}
\caption{Rotational history of PSR~B1800$-$21 with timing and frequency residuals
presented in upper and lower panels, respectively. Errors are 
smaller than size of the points in the upper panel and are typically 0.2 ms}
\label{fig6}
\vspace{2.0in}
\end{figure}

\subsection{PSR~B1917+00 (J1919+0021, $\tau=2,600$~kyr)}
Recently, around MJD~50174, the pulsar PSR~B1917+00 suffered a
small glitch, with fractional frequency increase $\Delta\nu_0/\nu_0 =
1.29 \times 10^{-9}$. There appears to be a small short-term relaxation after
the glitch.

\subsection{PSR~B1930+22 (J1932+2220, $\tau = 40$~kyr)}
A giant glitch occurred in this pulsar between MJD~50244 
and 50284. The fractional frequency increase, $\Delta\nu_0 / \nu_0 = 4457
\times 10^{-9}$, observed in this glitch is the second largest ever reported, 
marginally greater than the previous largest glitch in PSR B0355+54 
\cite{lyn87,sl96} and somewhat smaller than the recently reported glitch 
in PSR J1614$-$5047 (Wang et al. 2000). The 
frequency residuals and the frequency derivative for this glitch 
are shown in Figure \ref{fig9}. Removing the mean value of frequency 
from both pre-glitch and post-glitch data and expanding the 
frequency scale by a factor of 100 (Fig.~10b) 
reveals an almost linear relaxation of the rotation rate following a short 
term quasi-exponential relaxation. The  
exponential recovery can also be seen in the frequency derivative.

\subsection{PSR~B2255+58 (J2257+5909, $\tau=1,000$~kyr)}
PSR~B2255+58 exhibited a small glitch at MJD 49463. Timing and frequency
residuals corresponding to this glitch are 
presented in Fig. 11. This glitch is one of the smallest known with 
$\Delta\nu_0/\nu_0 = 0.92 \times 10^{-9}$.

\section{Discussion}
 
Shemar \& Lyne (1996) found 25 glitches in 10 pulsars after analysing
about 2500 years of pulsar rotation in Jodrell Bank pulsar timing data
base. Continuing this work we have found another 14 glitches in 9 
pulsars in a further 1000 years of pulsar rotation in the improved and
larger data base covering over about 7 years up to April 1997 (that is
around 50500 MJD). We found 6 pulsars glitching for the first time,
which increases the number of known glitching pulsars to 31. We also report 
new glitches in PSRs B1737$-$30, B1758$-$23 and B1800$-$21, which were already 
known as glitching pulsars. We report the smallest glitch ever
observed, in PSR~B1907$-$03 with $\Delta\nu_0/\nu_0 = 0.6 \times
10^{-9}$, and the second largest glitch, in PSR B1930+22 with 
$\Delta\nu_0/\nu_0 = 4457 \times 10^{-9}$. The fractional frequency 
increase in most of the glitches described in this paper is of the 
order of $10^{-9}$. Thus, we detected 4 glitches which were not 
reported by Shemar \& Lyne (1996) as they were below their threshold of 
$5 \times 10^{-9}$ in $\Delta\nu_0/\nu_0$.  We also found 10 more glitches 
in the epoch interval not covered by them. For those glitches newly found in 
the Shemar \& Lyne interval, the average, the maximum and the minimum values 
of $\Delta\nu_0/\nu_0$ were 1.9, 2.9 and 0.6, while for those detected since 
then, the values were 646.2, 4457 and 0.92, respectively (all in units of 
$10^{-9}$). This can be compared with corresponding values of the Shemar 
\& Lyne (1996) search: 1072, 4368 and 1.2, respectively. They detected
several giant glitches while our survey resulted mostly in the detection of 
rather small glitches.

\begin{figure*}
\setlength{\unitlength}{1cm}
\begin{picture}(16,16)
\put(-1.7,-1.5){\includegraphics{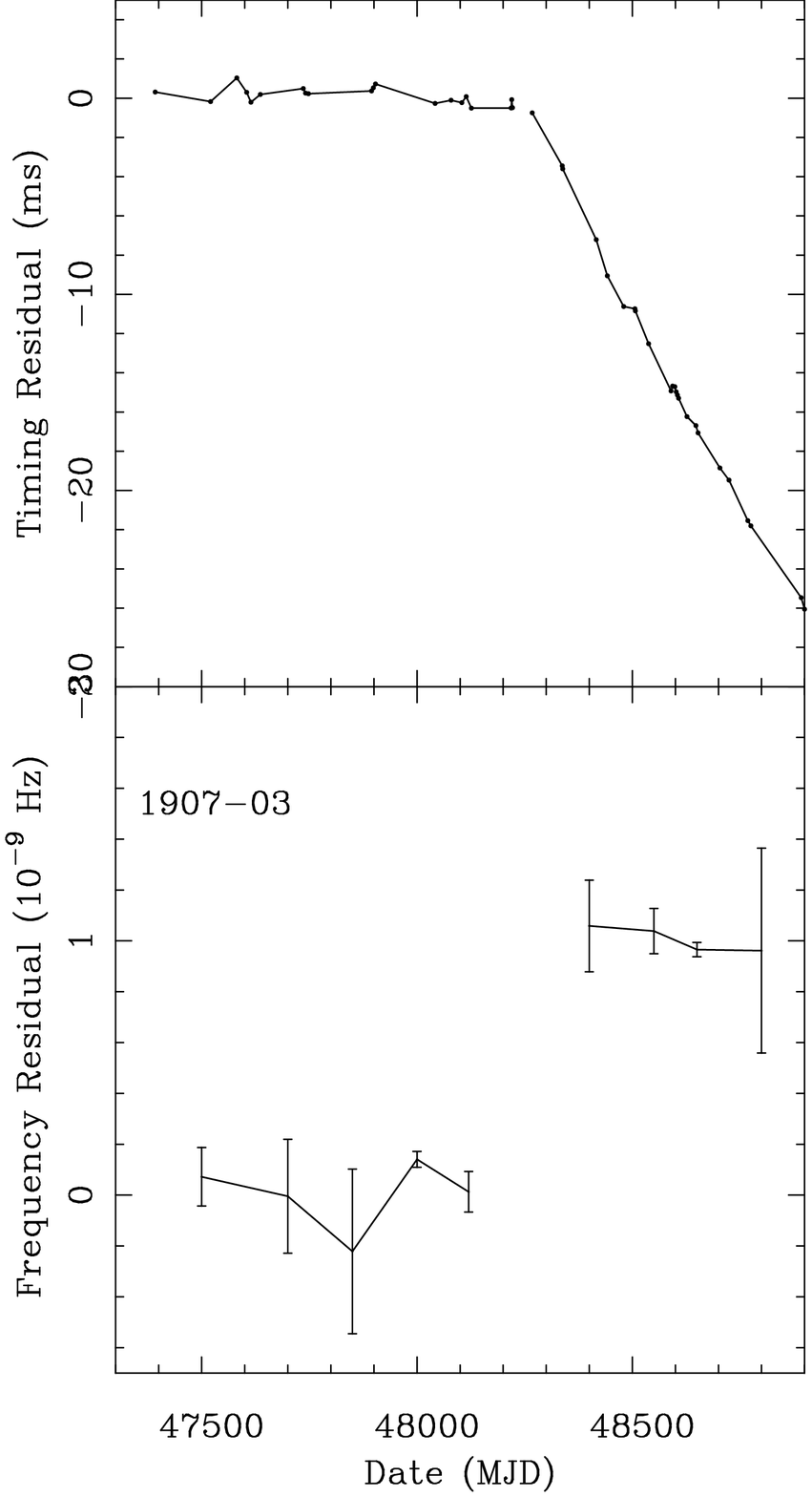}}
\put(7.0,-1.5){\includegraphics{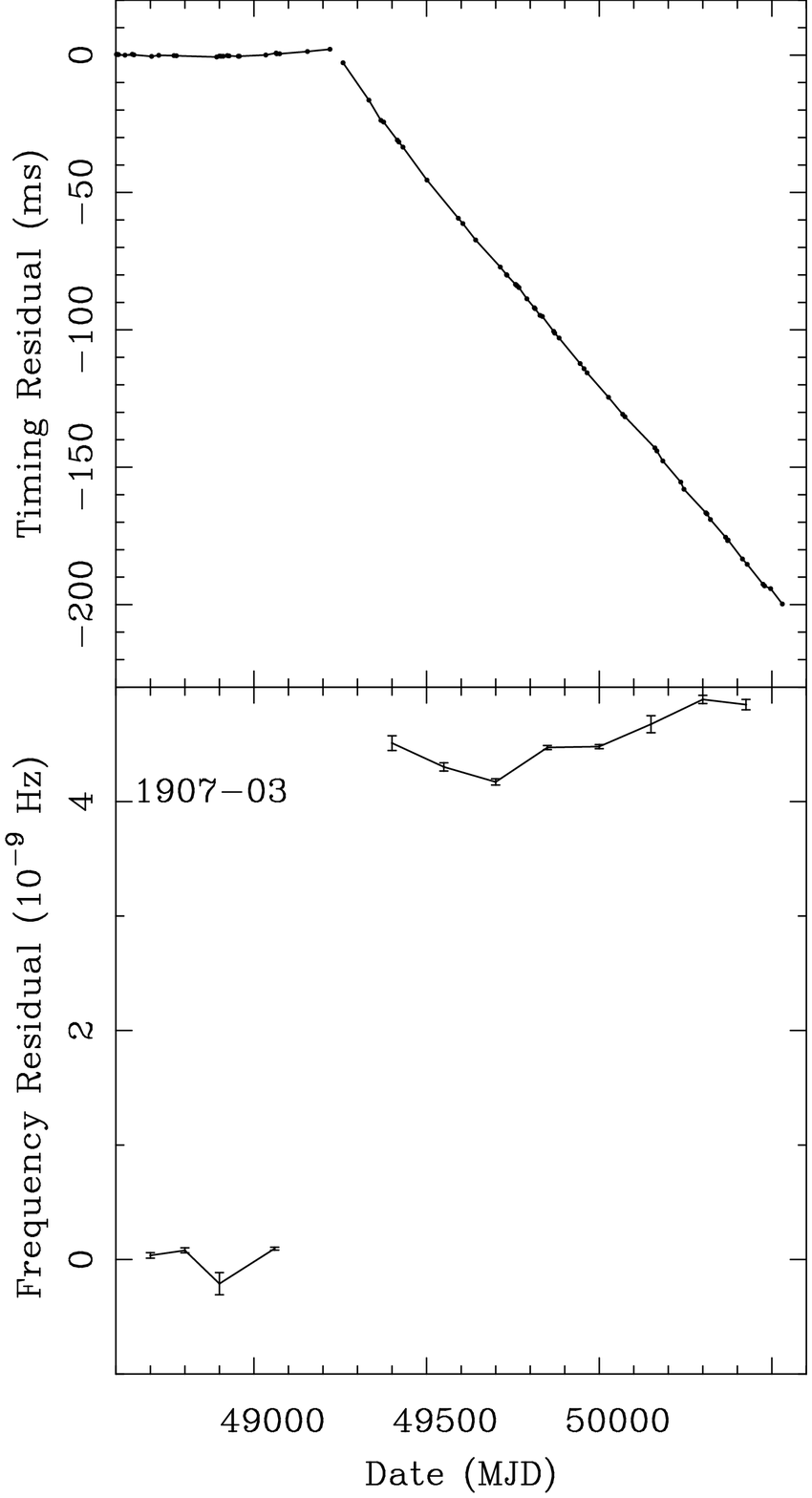}}
\end{picture}
\caption{Rotational history of PSR~B1907$-$03 with timing and
frequency residuals presented in upper and lower panels, respectively 
for the glitch around MJD 48241 (a) and that around MJD 49219 (b).
Errors are smaller than size of the points in the upper panel and 
are typically 0.3 ms and 0.2 ms respectively}
\label{fig7}
\vspace{2.0in}
\end{figure*}

\begin{figure}
\setlength{\unitlength}{1cm}
\begin{picture}(8,16)
\put(-1.7,-1.5){\includegraphics{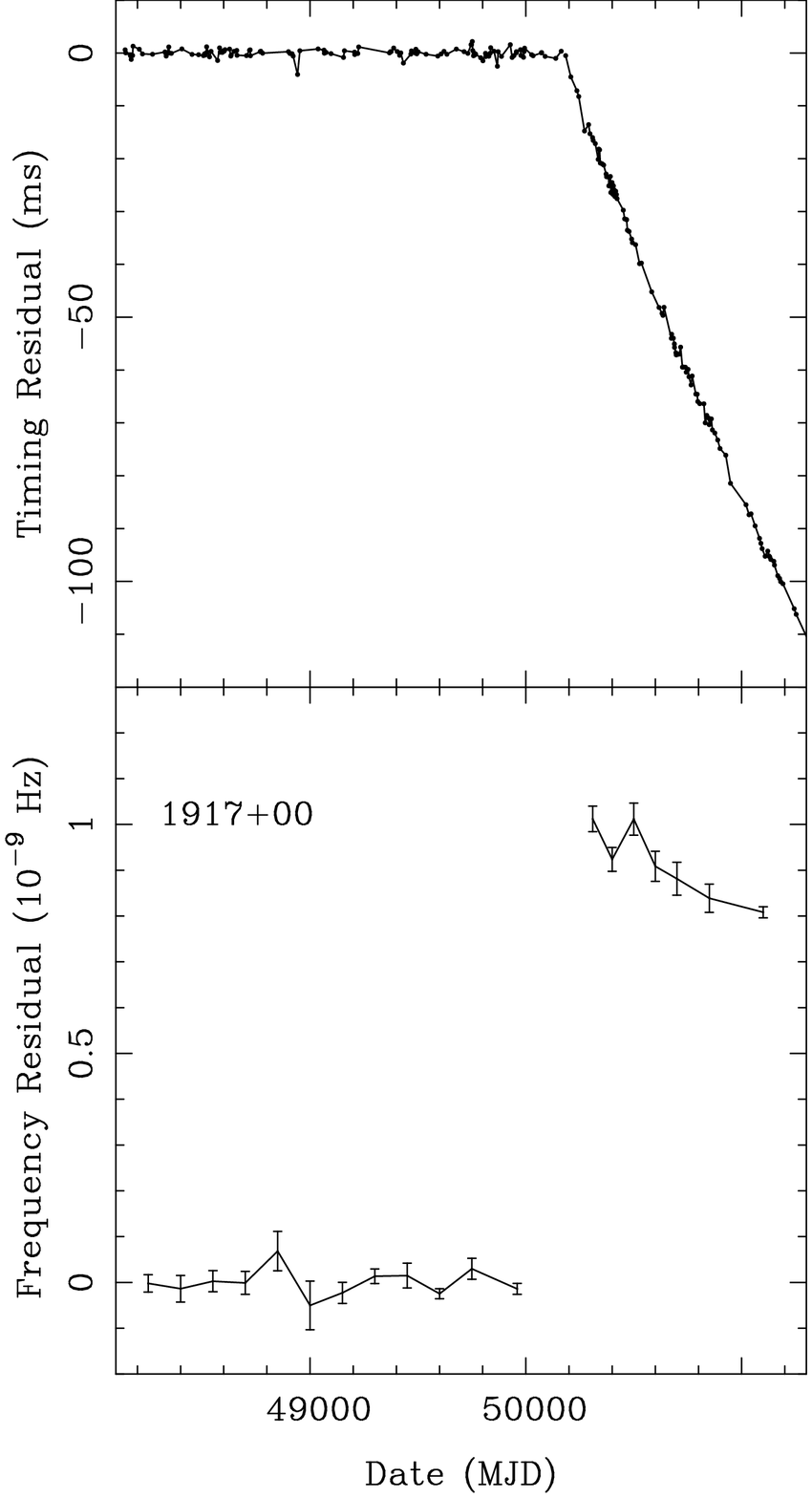}}
\end{picture}
\caption{Rotational history of PSR~B1917+00 with timing and frequency residuals
presented in upper and lower panels, respectively. Errors are smaller 
than size of the points in the upper panel and 
are typically 0.6 ms}

\label{fig8}
\end{figure}

\begin{figure}
\setlength{\unitlength}{1cm}
\begin{picture}(8,17)
\put(-1.7,-0.5){\includegraphics{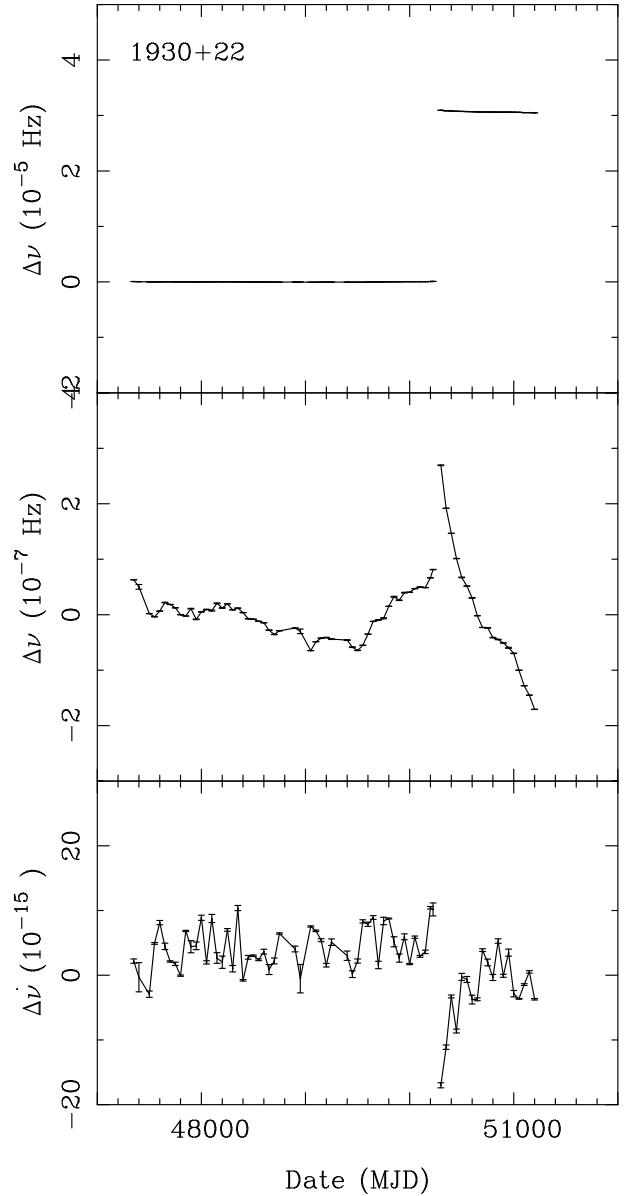}}
\end{picture}
\caption{The rotational history of PSR~B1930+22. a) The
frequency residuals $\Delta\nu$ relative to a simple slow--down model
involving the frequency and a constant value of its first
derivative. b) As for (a) but with the mean frequency in each 
interval subtracted and the vertical scale expanded by a 
factor of 100. c) The frequency first derivative with a constant value of 
-2760.0 subtracted.}
\label{fig9}
\end{figure}

The detection threshold of the present search for glitches varies
significantly from one pulsar to another, depending upon the accuracy
of the times-of-arrival, the density of the observations and the
presence of timing noise intrinsic to the pulsar.  We have conducted a
number of simulations on a few dozen representative pulsars which had
not glitched in our data set. The effects of a number of glitches of different
magnitude in turn were introduced into the arrival time data of a
pulsar.  These data were then inspected using the glitch detection
procedure described in section 2.  As a result of these tests, we
estimate that at least 90\% of glitches with magnitude $2 \times
10^{-9}$ have been detected.  This is a factor of about 2.5 smaller
than the threshold of Shemar \& Lyne (1996).  The detection of several
glitches below the threshold of those authors indicates that the
frequency of occurrence of small glitches does not decrease for
smaller sizes of glitch (e.g. Lyne {\it et al.} 2000).

\begin{figure}
\setlength{\unitlength}{1cm}
\begin{picture}(8,16)
\put(-1.7,-0.5){\includegraphics{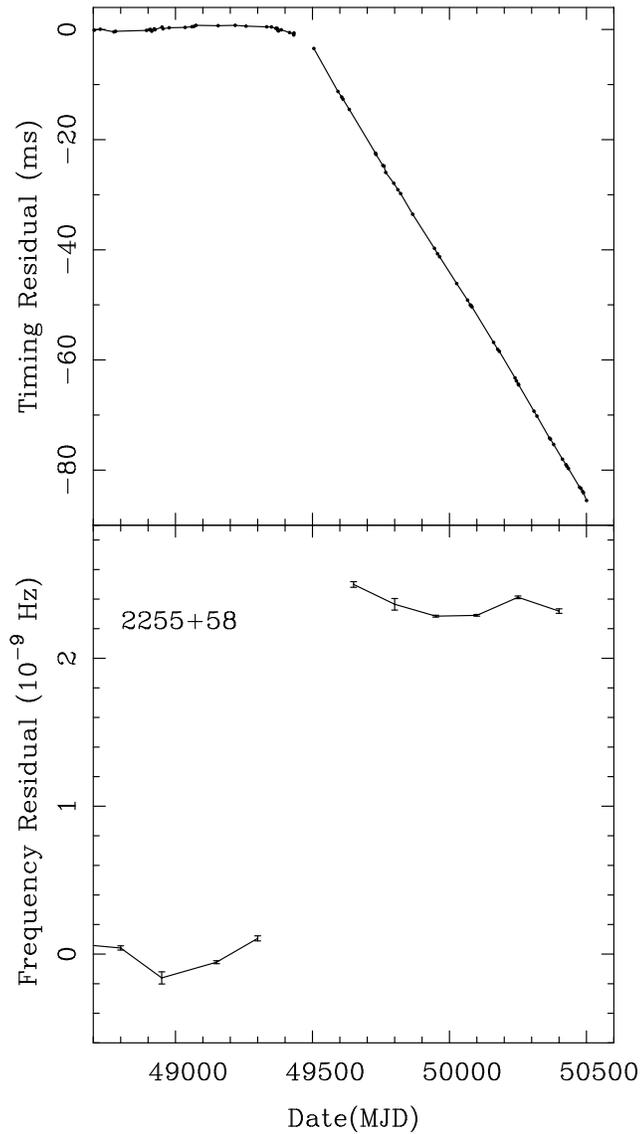}}
\end{picture}
\caption{Rotational history of PSR~B2255+58 with timing and frequency 
residuals presented in upper and lower panels, respectively for the 
glitch around MJD 49463. Errors are smaller than size of the points in 
the upper panel and are typically 0.04 ms }
\label{fig10}
\end{figure}

We confirm the observation of Shemar \& Lyne (1996) and Lyne {\it et
al.} (2000) that the dominant
effect of glitches, particularly the smaller ones, is a sudden
increase in rotational frequency with very little or no recovery. The
age range of the glitching pulsars is broad, from 16 thousand to 4.4 million
years, including PSR~B1758$-$03, which is now the second oldest
glitching pulsar.

\vspace {.5cm} \ni {\bf Acknowledgements} This work is supported in
part by the KBN Grant 2~P03D~015~12 of the Polish State Committee for
Scientific Research.

\bibliographystyle{mn}
\bibliography{modrefs,psrrefs,crossrefs}

\clearpage

\end{document}